\documentclass[12pt]{article}
\usepackage{graphicx}
\usepackage{bm}
\usepackage{epsfig}
\usepackage{amsmath}
\usepackage{amssymb}
\usepackage{color}
\usepackage[numbers]{natbib}
\usepackage{soul}

\newcommand{\beq}{\begin{equation}}
\newcommand{\eeq}{\end{equation}}
\newcommand{\beqa}{\begin{eqnarray}}
\newcommand{\eeqa}{\end{eqnarray}}

\newcommand{\fracd}[2]{\frac{d #1}{d #2}} 
\newcommand{\fracpd}[2]{\frac{\partial #1}{\partial #2}} 
\newcommand{\g}{g_{\mu\nu}}

\newcommand{\fracppd}[2]{\frac{\partial^2 #1}{\partial #2^2}}  
\newcommand{\gaa}{g_{00}}

\newcommand{\gdd}{g_{33}}
\newcommand{\gad}{g_{03}}

\begin{document}
\title{Experimental  tests of  pseudo-complex General Relativity}
\author{T. Sch\"onenbach$^1$, G. Caspar$^1$, Peter O. Hess$^{1,2}$,\\
Thomas Boller$^3$, Andreas M\"uller$^4$, Mirko Sch\"afer$^1$
and Walter Greiner$^{1}$ \\
{\small\it $^1$Frankfurt Institute for Advanced Studies, Johann Wolfgang Goethe Universit\"at,} \\
{\small\it Ruth-Moufang-Str. 1, 60438 Frankfurt am Main, Germany} \\
{\small\it
$^2$Instituto de Ciencias Nucleares, UNAM, Circuito Exterior, C.U.,}\\ 
{\small\it A.P. 70-543, 04510 M\'exico D.F., Mexico} \\
{\small\it $^3$Max-Planck Institute for Extraterrestrial Physics,} \\
{\small\it  Giessenbachstrasse, 85748 Garching} \\
{\small\it $^4$Excellence Cluster Universe, TU M\"unchen,} \\
{\small\it Boltzmannstrasse 2, 85748 Garching}}

\maketitle

\begin{abstract}
 Based on previous publications exploring pseudo-complex General Relativity (pc-GR) we 
present
a selection of observable consequences of  pc-GR and possible ways to experimentally access them. Whenever possible we 
compare the results to Einstein's GR and differences
 are worked out in detail. We propose 
experimental tests  to check the predictions of pc-GR for the orbital frequency
of test particles, the gravitational redshift effect and the last stable orbit.  We will show that the orbital
frequency of test particles at a given radius in pc-GR is in general lower compared to standard GR. Also the effect of frame dragging is modified (weakened) in pc-GR.
Concerning the gravitational redshift of a radiation emitting object we find that it is also lower in pc-GR than in standard GR. Eventually the classical concept of a last stable orbit has to be modified in pc-GR.
\end{abstract}


\vskip 1cm

\section{Introduction}

The theory of General Relativity (GR) has up to now withstood all
experimental tests. Nevertheless, there are extreme situations in GR,
like the formation of a singularity at the center
of a black hole, where one expects that quantum effects set in. 
The existence of a black hole itself is another example for such an extreme situation
as there is a part of space which is excluded from
 the access of an external, even nearby, observer.
The event horizon, which for a non-rotating object
is predicted in the external Schwarzschild solution of GR at the Schwarzschild radius $r_S$, divides
space into two parts, one for $r<r_S$ and the other one for $r>r_S$. Though, an infalling observer can access the internal region $r < r_S$, this part of space is inaccessible for an outside observer at a \textit{fixed} distance. Such an observer cannot extract information from the inside region while staying outside. This is one reason to search for possible extensions
to GR which avoid the appearance of an event horizon, making the internal region accessible for an external observer at a \textit{fixed} distance. Such a theory was
proposed by \cite{hess1} and modified by \cite{hess3}. An algebraic
extension of GR to pseudo-complex (pc) variables was proposed, called
pseudo-complex General Relativity (pc-GR). 
(In \cite{mann} it was shown that the only possible extension of GR
to different coordinates, avoiding so called
`ghost solutions', is the pseudo-complex extension.)
One of the important consequences
of pc-GR theory is the presence of an energy-momentum tensor
in the Einstein equation, which corresponds
to the distribution of a field
with repulsive properties.  It is something similar to dark energy.
This energy accumulates at mass distributions and increases when the mass density
gets larger.
One has to distinguish this effect from the 
one that a cosmological
constant $\Lambda$ has. 
The term in the metric containing the cosmological constant
is proportional to $r^2$ and thus increases
for greater distances. Therefore it is relevant 
on cosmological scales. In contrast, the gravitational repulsion in pc-GR can 
also get stronger
for smaller distances. This then has the effect that for very
large masses the gravitational collapse is stopped and
something what we call a `gray star'
is formed instead of
a black hole. The pc-GR theory
does not predict an
 event horizon, and thus 
the space with $r<r_S$
is accessible for an external observer.
In \cite{hess3} the pc-GR solutions for a non-rotating gray star (Schwarzschild),
a charged gray star (Reissner-Nordstr\"om) and a rotating gray star
(Kerr) were discussed. In \cite{hess2} the theory was applied to the
Robertson-Walker model of the universe, encountering new solutions,
for example, a finite or vanishing acceleration of the Universe for very large times.

It is important to investigate further cases, which can be verified
 by experiments, for example by observing
quasiperiodic oscillations due to orbital motions around 
a central large mass, see
\cite{Genzel2003,Aschenbach2004A,Iwasawa_2004}.
Very recently, a dense gas cloud has been detected falling into
the aggregation
 zone of the Galactic Center, see \cite{Gillessen_etal_2012}. 
The gas will be aggregated by the Galactic Center from 2013 onwards, and as the lower limit
of the absolute value of the black hole spin $|a|$ is larger than $0.5m$ this will allow for additional tests of
theories in the strong field limit.

Our pc-GR theory and standard GR 
make different predictions concerning the
angular velocity of test particles close to a large central mass. 
Experimental tests of those predictions
are discussed in this contribution.
The hope is that in the near future,
it will be possible
to detect phenomena in regions near the event horizon of standard GR.
The predictions of our
pc-General Relativity, which deviate from 
the standard General Relativity
theory then can be compared to experimental results.

This paper is organized as follows: In section 2 we give a very short
summary on pseudo-complex variables and pc-GR. In section 3 some 
experimental tests are proposed. We examine 1) the orbital
frequency of a particle around a gray star, 2) the radial dependence of the redshift, 
3) the structure of the effective potential for
the Schwarzschild and Kerr solutions, 
and finally a discussion of the
last stable orbit around a gray star. In all cases, we compare our results
to the the predictions of standard GR. Finally, in section 4 we present our conclusions.

In this article we use the signature (+,-,-,-) 
for the metric,
together with the definition of the
Riemann tensor which leads to the Einstein equations in the form $G_{\mu\nu} =\mathcal{R}_{\mu\nu} - \frac{1}{2} \g \mathcal{R} = - \frac{8 \pi \kappa}{c^4} T_{\mu\nu} $, with $c$ being the speed of light and $\kappa$ the gravitational constant.
The parameter $a$ containing
the angular momentum $J$ of a rotating compact massive object,
is defined as $a = \frac{- \kappa J}{m c^3}$ like in \cite{adler}.
Here $m$ is half the Schwarzschild radius of the considered object. It is related to the mass $M$ of the object
by $\frac{\kappa M}{c^2}=m \equiv \frac{r_s}{2}$. 
The parameter
$a$ has units of length and will be measured in multiples of $m$. 

\section{Pseudo-complex General Relativity}

In what follows,
a short review of the main properties of pc-GR
is given.
The pseudo-complex coordinates are $X^\mu = x^\mu + I\frac{l}{c} u^\mu$, with $x^\mu$
as the position and $u^\mu$ as a vector with units of a four-velocity.
The factor $l$ is introduced due to dimensional reasons. It represents
a minimal length parameter. $c$ is the speed of light. An important property is that a pseudo-holomorphic function
of pc-variables can be written as $F(X)=F_+(X_+) \sigma_+ + F_-(X_-)\sigma_-$
($X_\pm^\mu = x^\mu \pm \frac{l}{c}u^\mu$),
with $\sigma_\pm = \frac{1}{2}\left( 1 \pm I \right)$,
$\sigma_\pm^2 =\sigma_\pm$ and $\sigma_+ \sigma_- = 0$. Due to the last
relation, pc-variables have a
zero-divisor\footnote{For a commutative ring the zero-divisor
is the set of numbers $a,b $ which fulfill the relation
$a \cdot b = 0$ without being zero on their own.}
with $\sigma_\pm$ as its basis. Calculations can be done
independently in the $\sigma_+$ or $\sigma_-$ component. This 
plays a crucial role for the pc-extension of GR. For more details on
pc-variables and rules of calculations, see \cite{anton,kantor}.

In pc-GR the metric is pseudo-complex and written as
$g_{\mu\nu}(X) = g_{\mu\nu}^+(X_+) \sigma_+
+ g_{\mu\nu}^-(X_-) \sigma_-$. Due to the
fact that calculations can be performed independently in each
$\sigma_\pm$ component, a GR theory is constructed in each component. All rules
can be immediately copied from any text book on GR, e.g. \cite{misner,adler}.
An important ingredient is the modification of the variational principle,
first proposed in \cite{schuller0,schuller1}. The variation of the action is
now equal to a value within the zero divisor, i.e. $\delta S$ is either proportional to
$\sigma_+$ or $\sigma_-$, where for convenience the $\sigma_-$ is
chosen (to take $\sigma_+$ just gives an equivalent description where plus
and minus are interchanged). This modifies the Einstein equations which
are now
\beqa
G_{\mu\nu} = -\frac{8\pi\kappa}{c^2} T_{\mu\nu} \sigma_-
~~~.
\label{einstein-eq}
\eeqa
This represents a trivial extension to standard GR,
keeping the beautiful structure of the theory, only adding
an energy-momentum tensor, similar to theories which include matter.
The energy-momentum tensor describes the
contribution of a field which has repulsive properties and is
responsible for the halt of a gravitational collapse. Finally, the metric
$g_{\mu\nu}(X)$, obtained in
an actual case,
is projected to
$g_{\mu\nu}(x)$, where the function is the same but the pc-variables are
substituted by the coordinates and the pc-parameters by their pseudo-real
components. This corresponds to 
neglecting the contributions
due to the
minimal length $l$, see \cite{hess3}. We will not repeat the details here, 
but rather refer the reader to the references \cite{hess1,hess3}. 

The form of the energy-momentum tensor might be explained by microscopic physics  
\cite{visser1,visser2,visser3}, where the vacuum fluctuations (Casimir effect)
are discussed in the presence of matter. The Schwarz\-schild metric is treated as 
a background and the influence
of the matter on the vacuum fluctuation is deduced,
using different approaches for the vacuum. A strongly
decreasing energy density
as a function of distance is found.
The same should be repeated taking into account
the coupling of the vacuum fluctuations with the metric.
We mention this work here, because it shows that our ansatz
has a possible microscopic origin.

The following solution of equation (\ref{einstein-eq}) is the pseudo-complex analogue
of the Kerr metric of standard GR 
\begin{align}
g^{\text{K}}_{00} &= \frac{ r^2 - 2m r  + a^2 \cos^2 \vartheta + \frac{B}{2r} }{r^2 + a^2 \cos^2\vartheta} \notag \\
g^{\text{K}}_{11} &= - \frac{r^2 + a^2 \cos^2 \vartheta}{r^2 - 2m r + a^2 +  \frac{B}{2r}  } \notag \\
g^{\text{K}}_{22} &= - r^2 - a^2 \cos^2 \vartheta  \notag \\
g^{\text{K}}_{33} &= - (r^2 +a^2 )\sin^2 \vartheta - \frac{a^2 \sin^4\vartheta \left(2m r -  \frac{B}{2r}  \right)}{r^2 + a^2 \cos^2 \vartheta}  \notag \\
g^{\text{K}}_{03} &= \frac{-a \sin^2 \vartheta ~ 2m r + a \frac{B}{2r}   \sin^2 \vartheta }{r^2 + a^2 \cos^2\vartheta}   \quad ,
\label{eq:kerrpseudo}
\end{align}
which reduces in the limit $a=0$ to the pseudo-complex analogue of the standard GR Schwarzschild solution
\begin{align}
g^{\text{S}}_{00} &= \left( 1 - \frac{2m}{r} + \frac{B}{2r^3} \right) \notag \\
g^{\text{S}}_{11} &= -\left( 1 - \frac{2m}{r} + \frac{B}{2r^3} \right)^{-1} \notag \\
g^{\text{S}}_{22} &= - r^2  \notag \\
g^{\text{S}}_{33} &= - r^2 \sin^2 \vartheta  \quad .
\label{eq:schwarzschildpseudo} 
\end{align}
Here the superscripts `S' and `K' refer to `Schwarzschild' and `Kerr', respectively. 
In the further discussions we 
omit these superscripts, 
as a distinction between Schwarzschild and Kerr metric can be made via
the parameter $a$, which is a measure of the rotation of the object, see
\cite{adler,misner}.

Requiring a positive $g_{00}$-component of the metric
 leads to a lower possible limit $B>B_{{\rm min}}= \frac{64}{27}{\rm m}^3$.
In deriving the metric, we have to make special assumptions about the
energy-momentum tensor. As discussed in \cite{hess3}, a fluid model
for the dark energy is used. The density is assumed to be proportional
to $1/r^5$, in order that the additional metric component behaves as
$1/r^3$. The reason for this lies in astronomical observations
in the solar system \cite{will} where contributions of the order $1/r^2$ are already
excluded. In general, for the additional metric contribution
one can use a power expansion in $1/r$, starting
with $1/r^3$. This includes more parameters and for that reason,
keeping the extended theory as simple as possible, we take for the moment
only the leading term. Thus, only one additional parameter is included,
as expected when a theory is generalized. The effects of changing
the parameter $B$ and the power in $1/r$ are discussed in the corresponding
places of this article.

The pc-GR theory also eliminates the singularity at $r=0$. There, the
metric component $g_{00}$ diverges in ordinary GR.
Therefore, the theory presents a different suggestion on how to avoid
the singularity at the center.
However, this is only
of academic interest, 
because the massive object ceases contraction below 
the position of the minimum of the effective potential, which is 
proportional to $g_{00}$. For radial distances smaller than the position
of the minimum of $g_{00}$ mass should be present, which has to
be taken into account, changing the metric. This is the subject of our
current investigation, important also for neutron stars. For $r$ near
the origin, quantum effects will have to be taken into account, but this is not considered in the current work. 
The singularity problem is {\it not the topic discussed in this contribution} but rather we investigate the
consequences of the disappearance of the event horizon as for example done by \cite{mazur}.
\section{Experimental tests}
\subsection{Radial dependence of the angular frequency $\omega(r)$}
\label{sec:angularfrequency}

The observation of the circular movement of matter around a large
central mass might give hints for deviations
of  pc-GR from Einstein's GR.
For example, the orbital frequency of a plasma cloud can be measured.
We expect a spectacular infall event of such a plasma cloud very soon 
\cite{Gillessen_etal_2012} allowing to test GR very close to a black hole candidate.
Indeed,
this is already possible, because flares are observed
at the Galactic center black hole candidate Sgr A*, see
\cite{Genzel2003,Aschenbach2004A}.

We approximate
such clouds by a massless particle and therefore  calculate first
the angular frequency $\omega(r)$ of a
point particle orbiting a large central mass $M$ on a stable geodesic track, see \cite{hess3}.
The following discussion is related to subsection 3.3, where the
existence of a last stable orbit 
is investigated.
In order to determine the angular velocity,
we proceed similar to \cite{hess3} and
follow the book of Adler et al.
\cite{adler}, using the Lagrange function
\begin{align}
L = g_{00}c^2 \dot{t}^2 + g_{11} \dot{r}^2 + g_{22} \dot{\vartheta}^2 + g_{33} \dot{\varphi}^2 + 2g_{03} c\dot{t}\dot{\varphi} = \frac{ds^2}{ds^2} = 1  \quad ,
\label{eq:lagrange}
\end{align}
where the dot represents  $\fracd{}{s}$
and $s$ is a curve parameter, e.g. the eigentime.
The variation of $L$ then yields the geodesic equations from which we will use the radial one:
\begin{align}
\frac{d}{ds} \left ( 2g_{11}\dot{r} \right ) &=  g'_{00} c^2\dot{t}^2 + g^\prime_{11} \dot{r}^2 + g'_{22} \dot{\vartheta}^2 + g'_{33} \dot{\varphi}^2 + 2 g'_{03} 
c\dot{t}\dot{\varphi} \quad .
\label{eq:radgeo1}
\end{align} 
Here the prime $'$ stands for the derivative $\fracpd{}{r}$. To simplify the calculations we 
 restrict our discussion to motions in the equatorial 
plane at a constant distance to the central object,
e.g. $r = r_0$, $\dot{r} = 0$, $\vartheta = \frac{\pi}{2}$
and $\dot{\vartheta} = 0$.
In this case,
equation (\ref{eq:radgeo1}) becomes
\begin{align}
0 &= g'_{00}(r_0) c^2\dot{t}^2 + g'_{33}(r_0) \omega^2\dot{t}^2 + 2g'_{03}(r_0) \omega c\dot{t}^2 \quad ,
\label{eq:radgeo2}
\end{align}
where we introduced the angular frequency $\omega = \fracd{\varphi}{t} =  \frac{\dot{\varphi}}{\dot{t}}$. 

Obviously equation (\ref{eq:radgeo2}) is a
quadratic equation in $\omega$ with the two solutions
\begin{align}
\omega_{\pm} = c\frac{- g'_{03} \pm \sqrt{\left (g'_{03}\right )^2 - g'_{00}g'_{33}}}{g'_{33}} \quad. \label{eq:omegaStable}
\end{align}
Inserting now equation (\ref{eq:kerrpseudo})
yields
\begin{align}
\label{Kreisfrequenz_neu}
\omega_{\pm} = c\frac{-a h(r)\pm\sqrt{2rh(r)}}{-2r +-a^2h(r)}=
\frac{c\sqrt{h(r)}}{-a\sqrt{h(r)}\mp \sqrt{2r}} \quad,
\end{align}
with
\begin{align}
h(r)=\frac{2m}{r^2}-\frac{3B}{2r^4}\quad.
\label{hr}
\end{align}
For $h(r)>0$, equation (\ref{Kreisfrequenz_neu})
 has two real solutions, one for co- and one for counter-rotation with
respect to the rotating central body. Note that $h(r)>0$
is equivalent to $r^2>(3B)/(4m)$, which gives
$r>(4/3)m$ for $B=B_{{\rm min}}$. The task now is to find out
which solution of equation (\ref{Kreisfrequenz_neu}) corresponds to which
physical motion.

The angular momentum of the central mass is given by $J$, which might
have positive or negative sign. As usual we assign a
positive $J$ to a counter-clockwise rotation
(i.e., mathematical positive sense).
For convenience we define
that the angular momentum for this rotation points towards the North pole of the central mass.
 Next we have to use the connection between the parameter $a$ and the angular momentum $J$ as given in
\cite{hess3,adler} by

\begin{equation}
a = \frac{- \kappa J}{m c^3} \quad.
\end{equation}
Here one has to be careful as there also exists the convention $a = +  \frac{\kappa J}{m c^3}$ as used for example in 
\cite{misner,muller2}.

We choose $J$ to be positive and accordingly $a$ to be negative.
Since $h(r)>0$, it always holds that $|\omega_{+}|>|\omega_{-}|$.
First consider $2r>a^2h(r)$, which is always the case for large $r$.
In this case $\omega_{+}$ is negative and $\omega_{-}$ is positive.
Consequently $\omega_{-}$ describes co-rotating orbits,
whereas $\omega_{+}$ describes counter-rotating orbits.
Now consider $2r<a^2 h(r)$, which occurs for small $r$.
In this case $\omega_{+}$ and $\omega_{-}$ are both positive,
so {\it the previous counter-rotating orbit has turned into a co-rotating orbit}.

\begin{figure} 
\begin{center}
\includegraphics[width=\columnwidth]{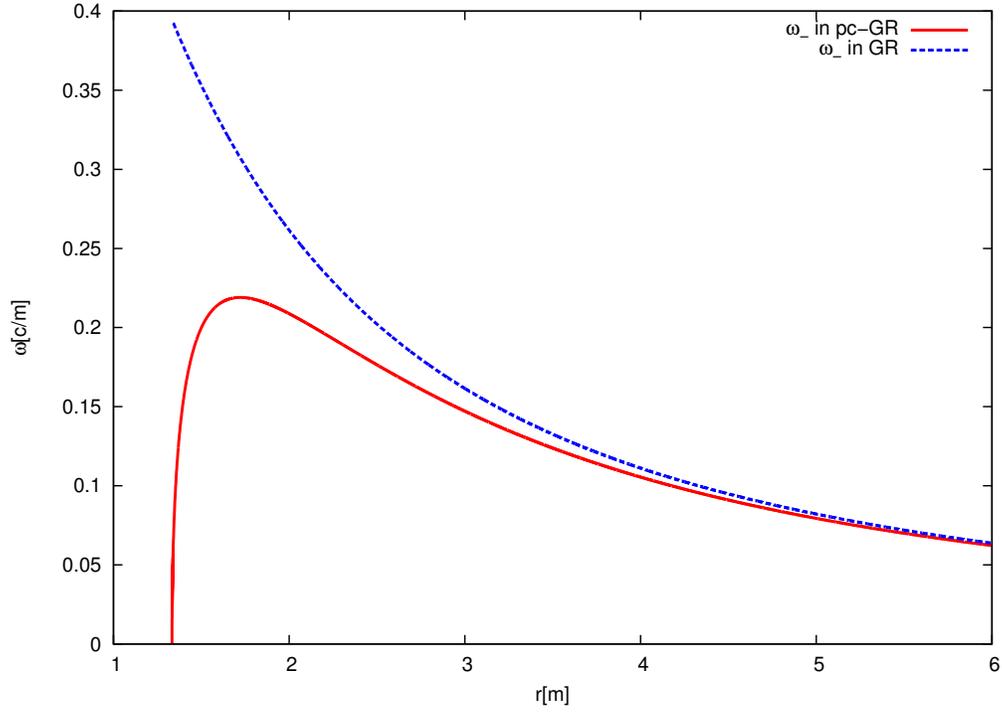} 
\begin{flushleft}
\caption{
Orbital frequency as a function of $r$, for stable geodesic prograde
(rotating in the same way as the central massive object)
circular motion.
The value $\omega$ = 0.219, for a gray star having mass $4 \times 10^6$ solar mass,
corresponds to about 9.4 minutes for a full circular orbit. This plot is made with parameter values of
 $a= -0.995m$ and $B = \frac{64}{27}m^3$.
\label{prograde}}
\end{flushleft}
\end{center}
\end{figure}

\begin{figure} 
\begin{center}
\includegraphics[width=\columnwidth]{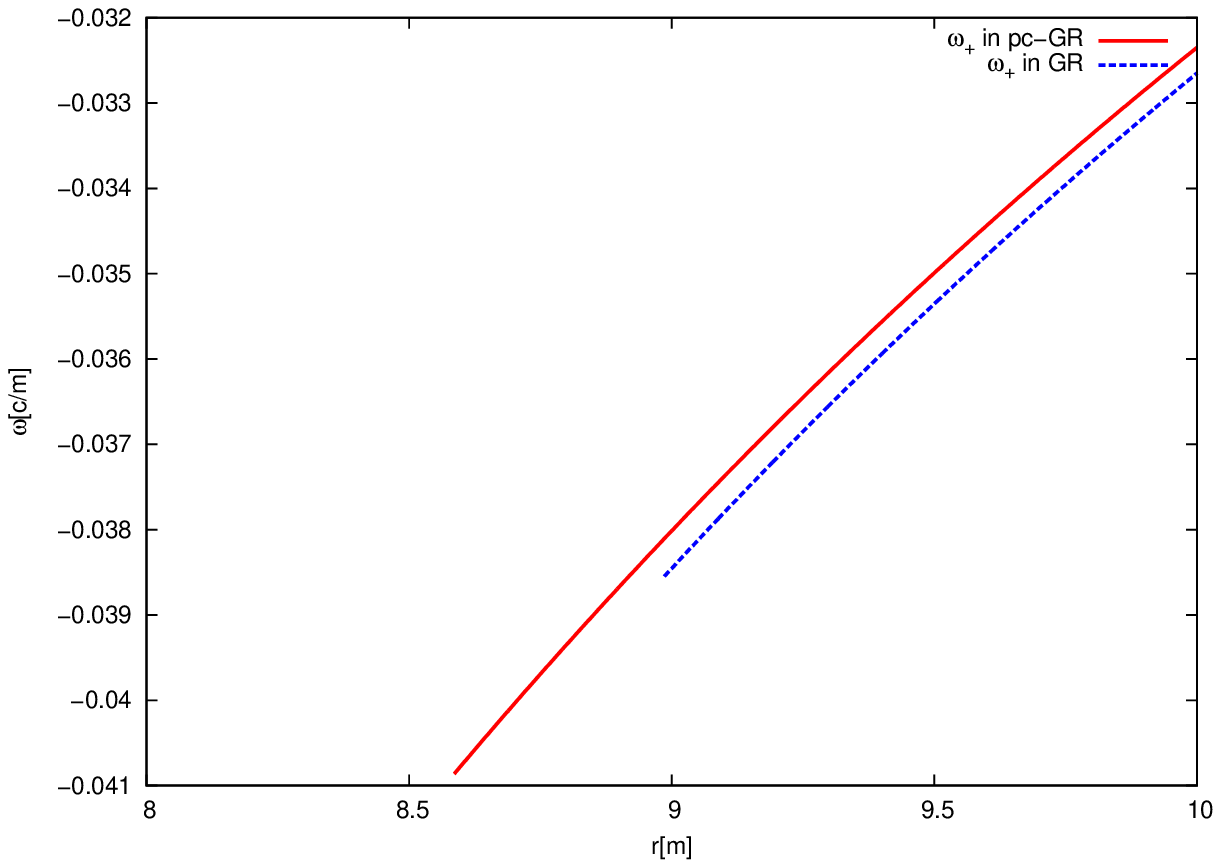} 
\begin{flushleft}
\caption{
Orbital frequency as a function of $r$, for retrograde
(rotating in the opposite way as the central massive object)
circular motion.
The parameters for this plot are $a= -0.995m$ and $B = \frac{64}{27}m^3$.
\label{retrograde}
}
\end{flushleft}
\end{center}
\end{figure}

In Fig.~\ref{prograde} the angular frequency $\omega_-$ of a mass in a
co-rotation
(prograde) orbit is plotted versus the radial distance,
with the rotational
parameter $a= -0.995m$.
 In Fig. \ref{retrograde} the same is shown for counter-rotation (retrograde orbit).
 In this case a
last stable orbit exists and the curve ends at a certain minimal value. The radial distance
is measured in units of $m$, and $\omega$ in units of $c/m$. For a mass of four million
times the mass of the sun \cite{Genzel2003}, corresponding to the object at the center of
our galaxy, a value of 0.219, at the maximum of $\omega$
= $2\pi \nu$ corresponds to a orbital period of $9.4$ minutes. 
In Fig.~\ref{prograde} we observe a maximum orbital frequency below $r=2$, i.e. below the Schwarzschild radius. Rewriting equation (\ref{Kreisfrequenz_neu}) as
\begin{equation}
\omega_{\pm}=\frac{c}{-a\mp\sqrt{\frac{2r}{h(r)}}}\quad,
\label{opm}
\end{equation}
we recognize
that the position of this maximum frequency is
{\it independent of the value of} $a$.
This is an important finding, because once a maximum in $\omega_-$ is observed, the position
of this maximum together with the mass will determine $B$.
It is given by the minimum of $r/h(r)$, which can be easily calculated to occur at $r=\sqrt{5B/4m}$. For $B=B_{{\rm min}}$ we obtain $r=\sqrt{80/27}m\approx 1.72m$, in perfect agreement with the numerical value in Fig. \ref{prograde}.
Starting below about 2 Schwarzschild radii,
differences between the standard GR
and the pc-GR appear, which become noticeable near $r=2m$
= $r_S$. In general, for a
particle it takes more time to circle the center
in pc-GR
than in GR. This difference
is expected to be observable
in the near future.
Care has to be taken though, because in standard GR the difference from pc-GR can masquerade as a different
(false) radius or a different value for $a$. Therefore, one has to combine this
observation with the measurement of the redshift, which also depends on
the radial distance. Only if the same radial distance is
obtained in both cases, is consistency achieved.
As will be shown in the last subsection, a stable orbit for prograde motion
always exists.
In the case of retrograde orbits we do not expect to see big differences as 
a last stable orbit exists at $r > 8m$ (for $a= -0.995m$). Below that value there
are no stable orbits anymore (see section \ref{sec:effpot} for details).
But in this region of space the difference between GR and pc-GR is
extremely small.

When instead of the $1/r^3$ dependence in the additional term in the metric
a higher power is considered, the qualitative behavior of the curve in Fig.~\ref{prograde} does not change, although the maximum is shifted. One can show that for a correction of order $1/r^{n}$ the maximum occurs at $r_{max}=\sqrt[n-1]{(n(n+2)B)/12m}$. For constant $B$ and $m$ this leads to increasing $r_{max}$, until around $n=10$ the value of $r_{max}$ decreases again and around $n=25$ reobtains its value as for $n=3$. Also $B_{{\rm min}}$ does depend on $n$ as $B_{{\rm min}}/2=(2m/n)^{n}(n-1)^{n-1}$.

Up to now we have only considered stable geodesic orbits. But there is also a possibility
to derive constraints to the orbital frequencies for more general orbits. Following
Misner et al. \cite{misner} we consider the demand that the line element
\begin{equation}
ds ^2 = g_{00} c^2 dt^2 + g_{11} dr^2 + g_{22} d\vartheta^2 + g_{33} d \varphi^2 + 2 g_{03} cdt d\varphi
\end{equation}
is positive (The difference is that
up to now the Lagrangian (\ref{eq:lagrange}) was varied
in order to obtain the geodesic equation, while
we consider now the general (non-geodesic) case.).
Again we restrict the calculations to circular ($dr = 0$) motions
which lie in the equatorial plane ($d\vartheta = 0 ~,~ \vartheta = \pi/2$).
The limiting case ($ds^2 = 0$) corresponding to a circularly rotating photon is then given by
\begin{equation}
g_{00} c^2 dt^2 + g_{33} \omega^2 dt^2  + 2 g_{03} c \omega dt^2 = 0 \quad , \label{eq:bedingungands}
\end{equation}
where we denoted
again $\omega = \fracd{\varphi}{t}$.
Equation (\ref{eq:bedingungands}) is a quadratic equation in
$\omega$ with the following solutions
\begin{equation}
\bar{\omega}_{\pm} =  c \frac{-g_{03}\pm  \sqrt{\left(g_{03}\right)^2 - 
g_{00}g_{33}}}{g_{33}}  \quad. \label{eq:omegaconstraint}
\end{equation}
Remarkably this is formally the same as equation (\ref{eq:omegaStable})
{\it but the metric coefficients appear instead of their derivatives}.
Inserting  equation (\ref{eq:kerrpseudo}) we get
\begin{align}
\bar{\omega}_{\pm}=c\frac{af(r)\pm\sqrt{D}}{ -\left(r^2+a^2\right) - a^2f(r)}\quad,
\label{omegaconstraint_neu}
\end{align}
with
\begin{align}
f(r)&=\frac{2m}{r}-\frac{B}{2r^3} \notag\\
D&= r^2+a^2-2mr+\frac{B}{2r} \quad .
\end{align}
In \cite{hess3} we have
shown that for $B\geq B_\text{min}= (4/3)^3m^3$ it holds $D\geq 0$.
Therefore, 
$\bar{\omega}_{\pm}$ has always two real solutions. It also holds
$f(r)\geq 0$ for $r\geq (3B)/(4m)$, that is $r\geq (4/3)m$ for $B=B_{min}$.
 Similar to the above discussion, when the power of $1/r$ is increased, the
$f(r)$ decreases, which in turn increases $\omega_\pm$.
This is modified by the $\sqrt{D}$ term.
When $B$ increases, also the $f(r)$ decreases and a similar effect is
observed, again modified by the $\sqrt{D}$ term in the denominator.
However, in general the structure remains the same.

We again choose $J$ positive and accordingly $a<0$.
In this case we have $\bar{\omega}_{-}> 0$ and
$|\bar{\omega}_{-}|\geq |\bar{\omega}_{+}|$.
Thus $\bar{\omega}_{-}$
describes the angular frequency of a co-rotating photon.
Since for large $r$ it holds $D\gg f(r)$, in this range
$\bar{\omega}_{+}$ is negative and corresponds to counter-rotation.
For smaller $r$ the term proportional to $1/r$ leads to an
increasing $f(r)$, and $\bar{\omega}_{+}$ might become zero.
In classical GR the sphere where $\bar{\omega}_{+}=0$ is called
the ergosphere. Since $g_{33}<0$,
it follows from equation (\ref{omegaconstraint_neu})
that the radius of the
ergosphere is given by the condition $g_{00}=0$. In pc-GR $g_{00}>0$
and thus
$\bar{\omega}_{+}$ is always negative. However, if $B=B_{{\rm min}}+\epsilon$
with $\epsilon\ll 1$,
$\bar{\omega}_{+}$ can become very close to zero at a certain radius.

\begin{figure} 
\begin{center}
\includegraphics[width=\columnwidth]{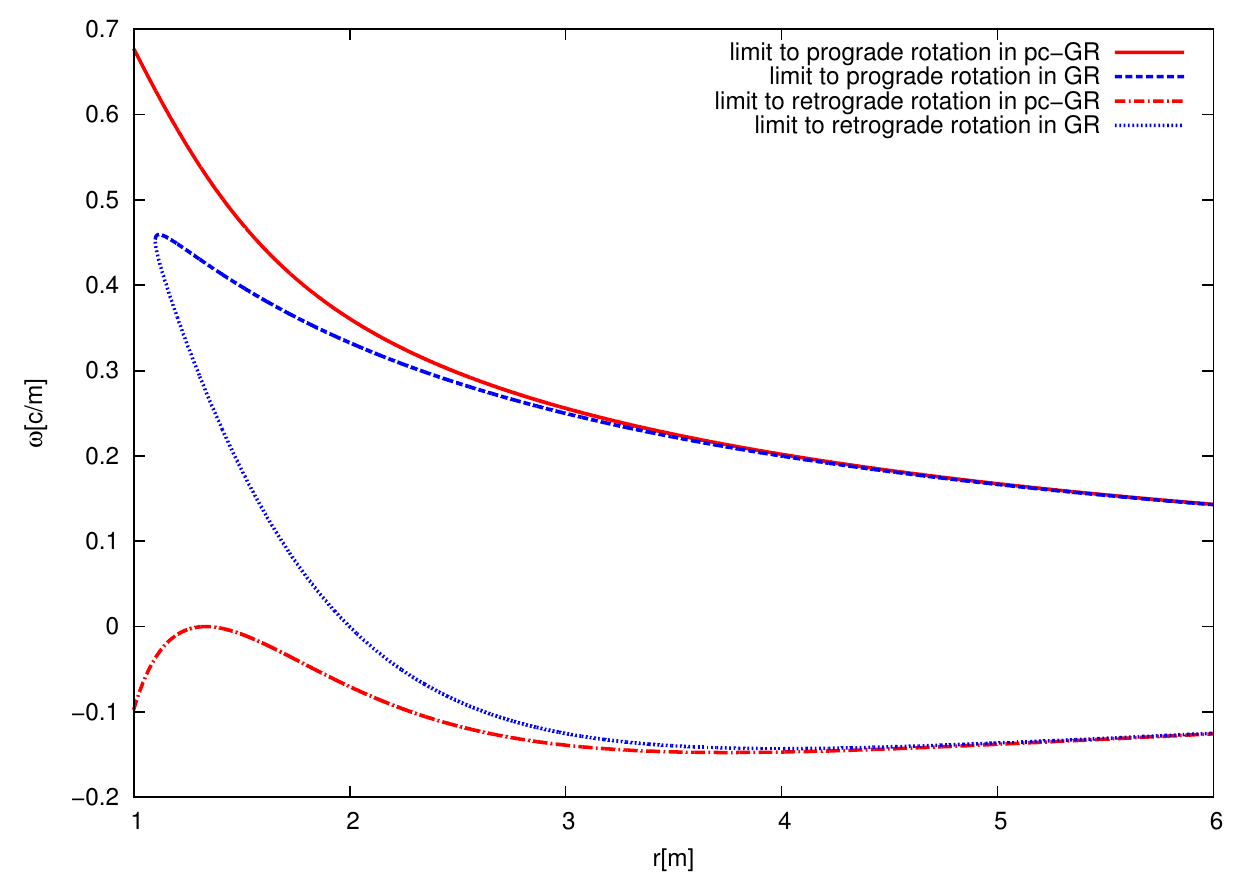} 
\begin{flushleft}
\caption{
Limits on the
orbital frequencies of circular motion in the equatorial plane.
The frequencies of particles moving on circular orbits must
lie between the limiting curves. The parameters
are $a= -0.995m$ and $B = \frac{64}{27}m^3$.}
\label{fig:omegalimits}
\end{flushleft}
\end{center}
\end{figure}

The results of equation (\ref{omegaconstraint_neu}) can be visualized quite nicely. 
In Fig.~\ref{fig:omegalimits} one can see the allowed range for circular movement (not necessarily on stable geodesics) 
compared between pc-GR and Einstein's GR.
The curves for pc-GR show a significantly different behavior as those for Einstein's GR. In standard GR there is a certain 
radius where the ergosphere begins. For smaller radii, particles have to co-rotate with the central mass, see 
\cite{adler,misner}. 
This behavior can be seen in Fig. \ref{fig:omegalimits} as the limiting curve for counter-rotating orbits changes its sign.
 The curves finally meet at the event horizon, where all particles have to rotate with the same frequency and in the same 
 direction as the black hole (frame-dragging).

This is different in pc-GR. Although the curve for counter-rotating orbits also approaches zero, there is no point 
where the two limiting curves coincide. In fact the point where the curve for counter-rotating orbits reaches zero also marks a 
maximum of this curve. 
It follows that although there is a frame-dragging effect in pc-GR
it is very small compared to  standard GR in the regime of strong gravitational fields. The allowed
range of values for circular motion is greater in  pc-GR than in standard GR. \\
For weak gravitational fields the effect of frame-dragging according to the predictions of \cite{lensethirring} has been confirmed experimentally, see \cite{ciufolini} and \cite{everitt}. But these predictions and measurements are not affected by the changes we consider here, as Lense and Thirring use a weak field approximation which stays the same in pc-GR and standard GR, see \cite{hess3}.

Now, we have firm prognoses at hand and
we suggest astronomical systems and methods which can be used to test the orbital frequencies.
Obvious testbeds are individual objects which orbit the black hole
candidate, e.g. something like planets in the solar system.
If we would observe
 a single star orbiting the compact mass we can simply apply 
GR versions of Kepler's laws.
The third law connects central mass, semi-major axis of the elliptical 
 orbit and orbital time, or orbital frequency respectively. If an astronomical observer is lucky enough to glimpse such a close stellar encounter,
  he could measure both orbital frequency and orbital radius, and put these measurements into the $\omega$-$r$ diagram. This would allow a direct test of
  whether the classical GR black hole or the pc-GR gray star model provides a better fit.
Slight modifications may arise when the $B/r^3$, as it appears
in the metric, is modified to $B/r^n$, ($n>3$), which produces changes as
discussed above.

Unfortunately, these stellar encounters are rare events. Typical rates for a massive black hole like the one in the center of our 
Milky Way are such that one star is swallowed every 10.000 years. Nevertheless, there are known cases 
where stellar tidal disruption 
events have been observed, see \cite{Gezari2006,Komossa2004}.
Indeed, if a star approaches a critical distance to a black hole \cite{Rees1988}, then
tidal forces disrupt the star and the stellar material becomes accreted onto the black hole. This
leads to an increase in luminosity. 
Such an event would be observed as an accretion burst accompanied by a luminosity outburst.
The so-called S stars are many stars observed in the infrared which orbit the 
galactic center black hole, see \cite{Gillessen_etal_2009}.
However, these stars are still too far away from the central mass to serve as probes to discriminate between GR and pc-GR.

Typically, the spatial resolution is not high enough to image details close to the 
Galactic center black hole. 
An alternative is to observe flares coming from that region. In fact, 
flares happen very frequently - typically once per 
day in the infrared at distances of about 30 Schwarzschild radii, see \cite{Genzel2003},
as well as in X-rays, see \cite{Aschenbach2004A}. 
But the nature of the flare emitter is still poorly understood.
A common model interprets the flare emission as modulated by relativistic Doppler effects and gravitational redshift, see references \cite{Genzel2003,Aschenbach2004A}. The flare emission oscillates and can be analyzed using power
spectra, i.e. Fourier transforms of the emission. Astronomers evaluate the flare frequencies by disentangling 
the prominent peaks in the power spectrum. Usually, one assumes a rotation of the flare at the
last stable orbit.
The shape of a spectral line emitted at the last stable orbit is 
directly related to the black hole spin, see \cite{Mueller2006}.
Therefore, one can deduce the black hole spin by matching the
spin-dependent last stable orbit, which is calculated theoretically, with the observed last stable orbit.
In fact, this is a common method to measure
the spin of a black hole.
But the last stable orbit is different if one applies the pc-GR theory, 
see section \ref{sec:effpot}. 
Hence, the whole analyses done in GR should be repeated in the pc-GR framework to search 
for a conclusive result. This is in addition motivated by the crucial point that state-of-the-art infrared flare observations of the Galactic 
center black hole hint for a spin parameter  $a < -0.5m$, which is not the same as that deduced from the X-ray flares, 
$a\sim -0.99m$.
Of course it should be the same black hole
(or gray star)! Such measurements will be improved significantly in the future
by using the GRAVITY experiment, see \cite{Eisenhauer_GRAVITY}, 
in the infrared and 
new generations of Athena-like X-ray observatories, see \cite{nandra_athena}.
A summary of future near-infrared and X-ray observations to test GR theories
in the strong field limit is given by \cite{Boller_Mueller_Makutsi}.

\clearpage 

\subsection{ Gravitational  redshift for Schwarzschild- and Kerr-type solutions}

\label{sec:redshift} 
The redshift of a spectral feature
carries valuable information and can be 
measured rather easily
by astronomers. \\
Let us consider a light ray,
which is sent by a
massless particle at rest in free space near a strong gravitational source to an observer at infinity.
To calculate the gravitational redshift 
we need the relation between the proper time and the coordinate time of the observer, which can 
be extracted from the line element
\begin{align}
ds^2 = c^2d\tau^2 &= g_{00}c^2 dt^2 + g_{11} dr^2 \notag \\
&\quad +  g_{22} d\vartheta^2 + g_{33} d\varphi^2 + 2g_{03} cdtd\varphi \quad .
\end{align}
Since the particle is considered at rest the equation simplifies to
\begin{align}
d\tau^2 = g_{00}dt^2 \quad .
\end{align}
In most cases we can assume, that the metric does neither change in the time between two wave peaks (here denoted by $\tau_0$
respectively 
t$_{obs}$), nor for the space between the particle and the observer while the ray is traveling. 
In this case the equation can be integrated and the result is obviously
\begin{align}
\tau_0 = \sqrt{g_{00}} t_{obs} \quad .
\end{align}

\begin{figure} 
\begin{center}
\includegraphics[width=\columnwidth]{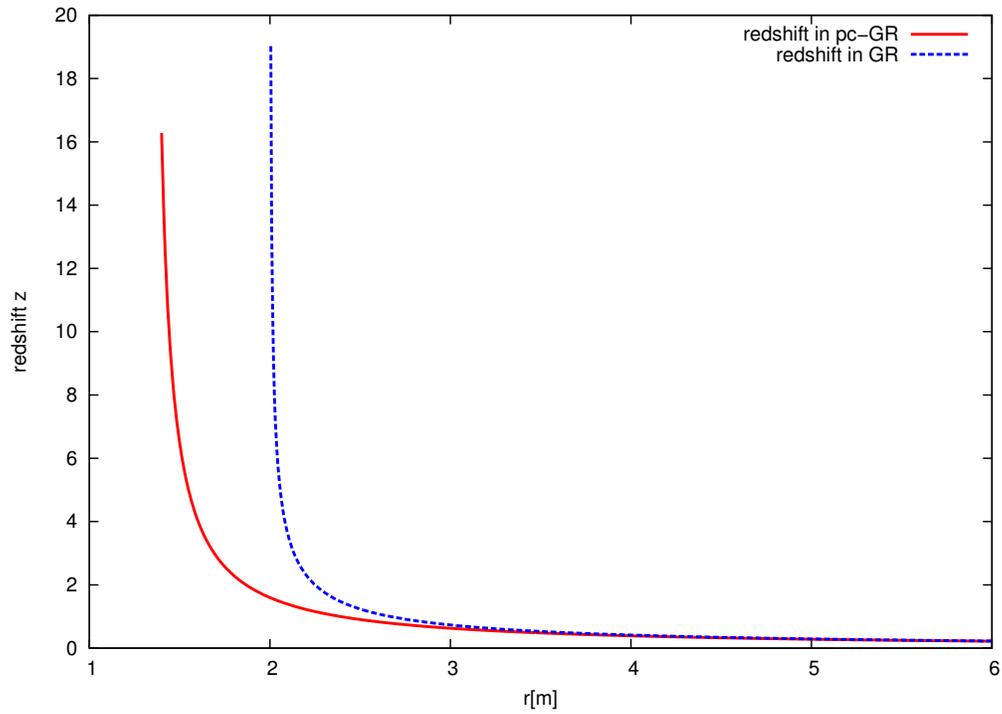} 
\begin{flushleft}
\caption{
Redshift of an emitter at the position $r$ in the outside field of a spherically symmetric, uncharged and static mass
(Schwarzschild metric) and also for the field at the equator of a rotating mass ($\vartheta = \frac{\pi}{2}$). $B$ is set to $\frac{64}{27} m^3$  and $a=0.995m$.
\label{redshift90}}
\end{flushleft}
\end{center}
\end{figure}  

Furthermore, since the times represent the interval between two wave peaks, they are the inverse of
the frequencies in their respective system. Hence the observed frequency differs from the emitted in the following way
\begin{align}
\nu_{obs} = \sqrt{g_{00}} \nu_0 \quad .
\end{align}
Following the convention in the literature, we define the parameter $z$ to obtain a measure for the redshift
\begin{align}
z := \frac{\nu_0 - \nu_{obs}}{\nu_{obs}} = \frac{1}{\sqrt{g_{00}}} - 1 \quad .
\end{align}
Using equation
(\ref{eq:schwarzschildpseudo}) we get for the 
pseudo-complex analogue of the Schwarzschild metric
\begin{align}
z = \frac{1}{\sqrt{1-\frac{2m}{r} + \frac{B}{2r^3}}} - 1 \quad ,
\end{align}
while the redshift in the pc-Kerr metric can be calculated by inserting
$g_{00}$ from equation
(\ref{eq:kerrpseudo})
\begin{align}
z = \frac{\sqrt{r^2 + a^2\cos^2(\vartheta)}}{\sqrt{r^2 - 2mr + a^2\cos^2(\vartheta) + \frac{B}{2r}}} - 1
~~~.
\end{align}
Note that the redshift at the equator ($\vartheta = \frac{\pi}{2}$) of the Kerr solution
is exactly equal to that in the Schwarzschild case.
This redshift is shown for the limiting case $B = \frac{64}{27} m^3$ in Fig.~\ref{redshift90}. One can clearly see, that between two and three
gravitational radii pc-GR differs noticeable from standard GR. The pc-GR theory predicts smaller redshifts for the same radius, so 
we would expect, to be able to look further inside. In this particular case the redshift still diverges at 
$r=\frac{4}{3}m$, but in reality we
expect a slightly larger B with the result that the redshift remains finite for the
whole space-time.
For a larger B, the curve for the redshift in pc-GR remains
finite in Fig.~\ref{redshift90}. In general, the redshift $z$ is decreased.
Also when the power in $1/r$ is increased, the value of $z$ decreases.
The main structure, however, remains which corresponds to smaller redshifts
in pc-GR, compared to GR, especially noticeable near the Schwarzschild
radius.
In conclusion, we predict that in pc-GR the surface of the central mass could in principle be visible. However it can be strongly
obscured due to the gravitational redshift.

\begin{figure} 
\begin{center}
\includegraphics[width=\columnwidth]{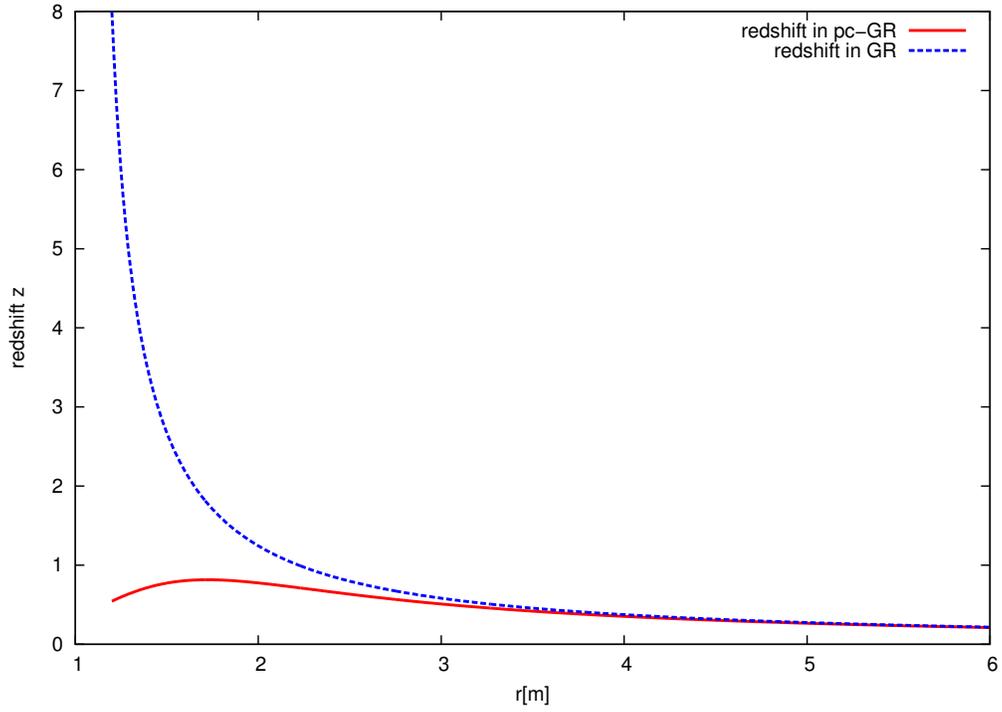} 
\begin{flushleft}
\caption{
Redshift for an emitter at the position $r$ in the outside field of
an axially symmetric, uncharged and rotating mass
(Kerr metric) at the poles (e.g. $\vartheta = 0$ or $\vartheta = \pi$). $B$ is again chosen to be $\frac{64}{27} m^3$ and $a=0.995m$.
\label{redshift0}}
\end{flushleft}
\end{center}
\end{figure} 

\begin{figure} 
\begin{center}
\includegraphics[width=\columnwidth]{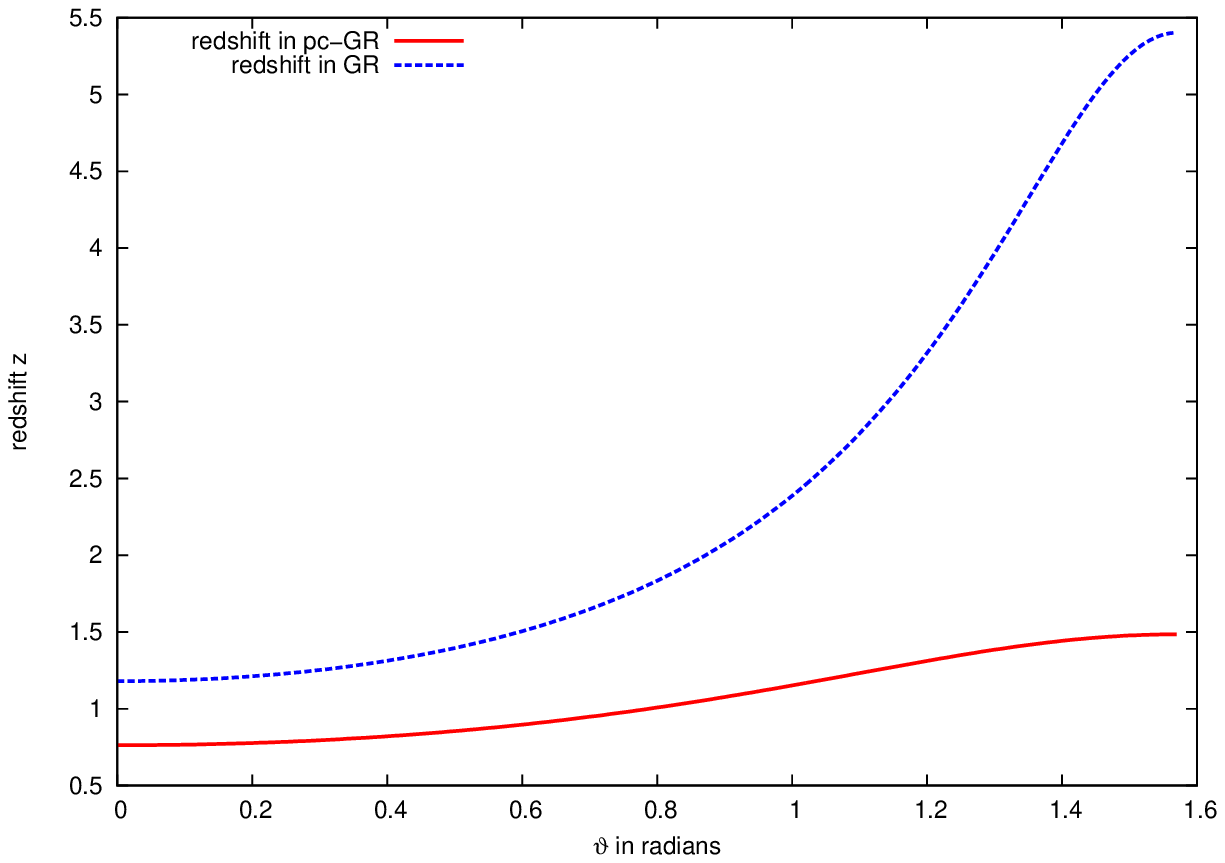} 
\begin{flushleft}
\caption{
Redshift for an emitter at the radius $r=2.05m$ in the outside field of
an axially symmetric, uncharged and rotating mass
(Kerr metric) for $\vartheta$ between $0$ and $\frac{\pi}{2}$. $B$ is equal to $\frac{64}{27} m^3$ and $a=0.995m$.
\label{redshiftTheta}}
\end{flushleft}
\end{center}
\end{figure} 

Nevertheless this result is not the whole picture for the Kerr solution,
since so far we have neglected the dependence
on the
inclination $\vartheta$. Compared to the previous result the other extremes are the poles, i.e. 
$\vartheta = 0$ or $\vartheta = \pi$. This case is shown in Fig. \ref{redshift0}. Here we see a completely different behavior,
since the maximal redshift is roughly 1 and the central object should be clearly visible.
This means that
the observation of a massive object from well separated angles 
could test our predictions.
 Figure \ref{redshiftTheta} shows the dependence of the redshift on the inclination $\vartheta$ for a given radius $r= 2.05m$. It thus gives an interpolation between the extremes shown in Fig. \ref{redshift90} and Fig. \ref{redshift0} for a certain radius. \\ 

The crucial difference between classical GR black holes and the
pc-GR gray  stars is that the pc-GR object
appears brighter for the same mass. In other words, the prominent
feature of a black hole, i.e. its blackness, is
is reduced in pc-GR.
A problem which might arise in this
context is that one cannot readily distinguish standard
black holes from pc-GR objects because one might mistake a pc-gray
star for a standard black hole by underestimating the mass. To prevent
this, astronomers should combine
various methods to weigh black holes (see e.g.
\cite{MuellerPoS} for a review on black hole mass and spin determination).

\begin{figure}
\begin{center}
\includegraphics[width=\columnwidth]{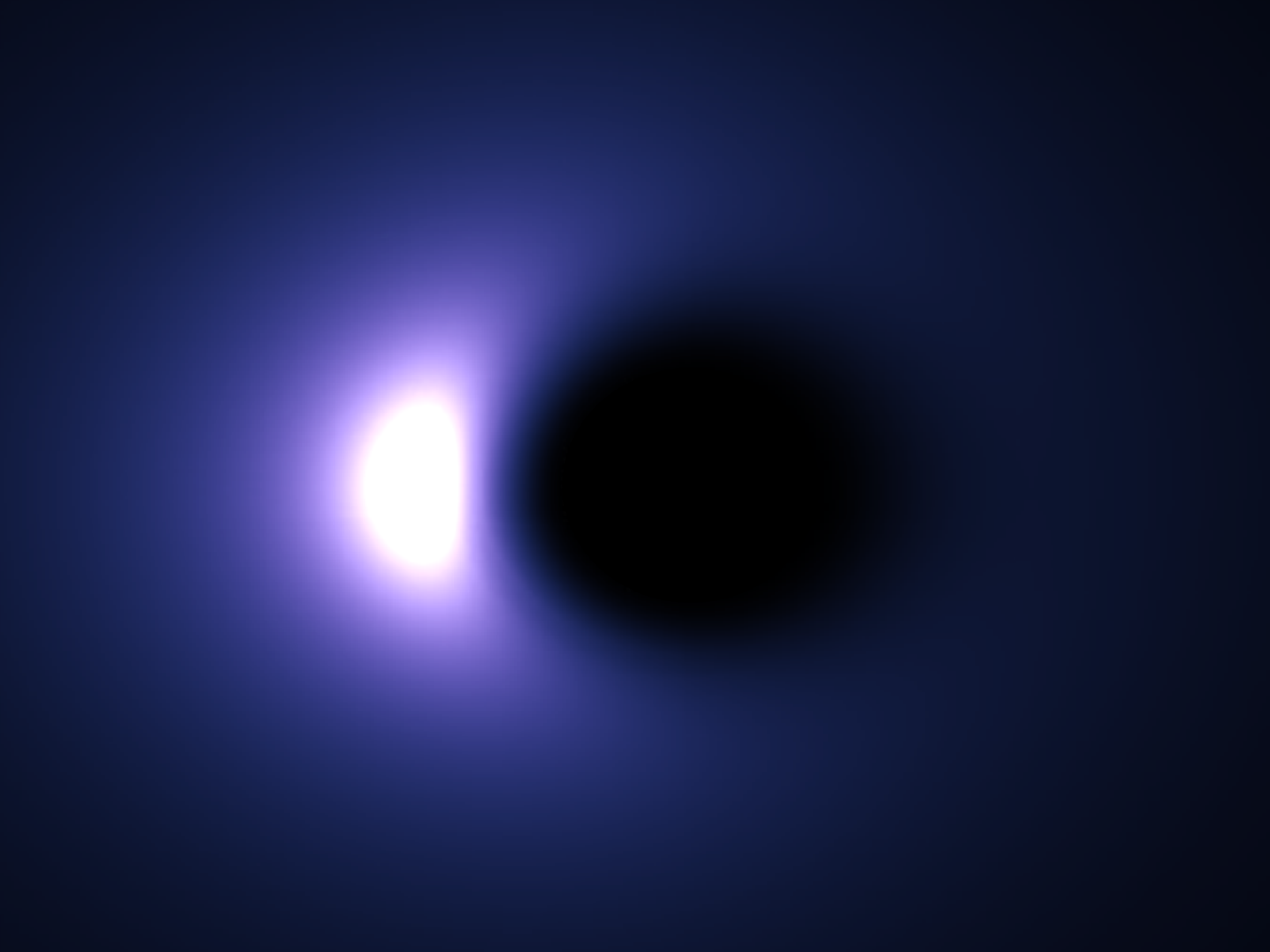} 
\begin{flushleft}
\caption{Luminous counter-clockwisely rotating accretion disk around an extremal Kerr black hole (standard GR), see 
\protect\cite{MuellerPoS}. The inclination angle
 is 40 degrees. The intensity is color-coded and grows from black (zero emission) over blue to white (maximum emission). On the left-hand side, the bright beaming feature due to Doppler blueshift is clearly visible. The black hole's horizon is seen in the middle of the image.
  \label{fig:raytracing}}
\end{flushleft}
\end{center} 
\end{figure}

An obvious method is to image the direct emission from a
 massive object and its immediate surroundings. For an illustration, 
see the GR ray tracing simulation in Fig. \ref{fig:raytracing}.

The event horizon is proportional to the black hole mass $M$, as
can be seen from the Schwarzschild radius $r_S = \frac{2 \kappa M}{c^2}$.
Therefore, astronomers could in principle weigh a black hole by measuring the 
observed black spot. 
Unfortunately, 
black holes are very
compact objects and the black hole candidates known so far subtend tiny angle s on the sky.
Only for two nearby massive
 black-hole candidates are black hole spot measurements currently feasible,
i.e. Sgr A* and M87 
in the Virgo cluster.
Concerning stellar-mass black hole candidates there seems to be no chance to succeed in 
applying this techniques - at least with the state-of-the-art generation of detectors.
We show this by evaluating the apparent size, i.e. 
the size of the black hole event horizon as viewed from the distance.
The angular scale for imaging the event horizons amounts to 
micro arc-seconds only.
Such a tiny size at the sky can only be imaged by using 
sophisticated techniques called
{\it Very Long Baseline Interferometry} (VLBI) in radio astronomy.
For astrophysics, an interferometer is an array of telescopes observing 
together astronomical objects.
The advantage is that the resolution scales with the distance of the 
individual telescopes. For VLBI the resolution corresponds to
a single telescope which was thousands of kilometers in diameter. 
Indeed, radio astronomers
have the best chance to image a black hole directly. 
They use these methods also to image the jet launching 
area { in the vicinity of black holes in the
heart of active
galaxies, see \cite{Krichbaum}.
The jet launching area is most probably the inner part of the accretion disc
surrounding the black hole, which is ejected perpendicular to the plane of the accretion disc
due to magnetic forces.
Until these imaging techniques are feasible - probably in the next few years - 
we have to wait for 
direct tests of the redshift effect in the the pc-GR theory. 
Infrared astronomers hope to detect GR effects, e.g. $\beta^2$ effects on stellar 
orbits around the Centre of our Galaxy, Lense-Thirring precession of the orbiting stars as well as flares from
the last stable orbit, with the powerful GRAVITY instrument mounted at 
ESO's Very Large Telescope (VLT) in Chile.
The VLT is the world's most advanced optical instrument, 
consisting of four Unit Telescopes with main mirrors of 8.2m diameter and 
four movable 1.8m diameter Auxiliary Telescopes. 
Astronomers can test the redshift effect also if the spatial resolution is not 
high enough. They do this using spectral methods. 
The spectral techniques are based on the canonical `lamppost' model, see
\cite{GeorgeFabian1991}. A hot corona surrounds the central black hole
and emits a power law spectrum. The X-rays from the hot corona will also
be reflected off the accretion disc resulting in the emission of fluorescent
lines. As shown by \cite{Mueller2006} the shape of the emitted lines
gives information about the distance of the line to the black hole, the
amount of gravitational redshift and the inclination of the accretion disc.
In the rest frame the line is usually well approximated by a Gaussian profile. 
However, in the observer's 
frame various effects influence the line profile,
namely the
relativistic Doppler effect and the
gravitational 
redshift effect (see e.g. \cite{MuellerCamenzind}). 
Hence, the evidence for gravitational redshift can be 
extracted 
from the red wing of the line. Typically, the closer the line 
emitting region extends 
towards the black hole, the greater the reduction in line emission at 
the red wing and the greater the reduction in the line flux. 
This idea was exploited to measure black hole spin because the last 
stable orbit of a 
standard Kerr black hole is significantly 
smaller than for a Schwarzschild black hole, see \cite{Tanaka1995}.
The fluorescent lines in the X-ray range are well suited
for determining basic parameters of the X-ray
emitting regions very close to the central black hole.
The most prominent fluorescent line is the strong iron K$\alpha$ line, visible
at 6.4 keV rest frame energy. In a similar energy range there is also a
weaker iron L line. These lines are not visible in all black hole candidates, but
can be seen for those cases
where a cold geometrically 
thin and optically thick accretion disk is irradiated by hard X-ray radiation 
from a hot nearby source.

A full solution would require putting the metrics of a pc-Schwarz\-schild and a pc-Kerr
gray star into a ray tracing code which solves the geodesic equation for a number of light rays (i.e. the null geodesic equation). 
However, this complex task lies beyond the scope of this paper.

Figure~\ref{fig:pro}
shows that a pc-GR Schwarzschild
object could mimic a standard Kerr black hole because the last stable orbit
is closer to the event horizon in the pc-GR case. This could be measured using iron K lines because fits deliver 
constraints on the inner disk edges. However, one now has a new situation with maybe completely different spin parameters. 
Ideally astronomers combine 
a vareity of
spin determination methods to constrain robustly the true spin value. 

\clearpage

\subsection{Effective  potentials and circular orbits}
\label{sec:effpot}

In this section we derive the effective potential for the radial motion of a geodesic in the equatorial plane,
e.g. $\vartheta = \pi/2$. For this case the variation of the Lagrangian (\ref{eq:lagrange}) leads to the following geodesic equations for $t$ and $\varphi$:
\begin{align}
0 &= \frac{d}{ds} \fracpd{L}{\dot{t}} = \frac{d}{ds} \left ( 2g_{00} c^2\dot{t} + 2g_{03} c\dot{\varphi} \right ) \notag \\
0 &= \frac{d}{ds} \fracpd{L}{\dot{\varphi}} = \frac{d}{ds} \left ( 2g_{33} \dot{\varphi} + 2g_{03} c\dot{t} \right ) \quad ,
\end{align}
which we write as
\begin{align}
g_{00}c\dot{t} + g_{03} \dot{\varphi} &=:\tilde{E} \notag \\
g_{33} \dot{\varphi} + g_{03} c\dot{t} &=: -\tilde{L} \quad,
\end{align}
where $\tilde{E}$ and $\tilde{L}$ can be identified with the energy $\tilde{E}=E/\mu c^2$ and the angular 
momentum $\tilde{L}=L/\mu c$ per mass $\mu$ of a test
particle, see \cite[p.~266ff.]{adler}.
An elementary rearrangement yields
\begin{align}
D  c\dot{t} &=  -g_{03} \tilde{L} - g_{33} \tilde{E}  \notag \\
D \dot{\varphi}  &= g_{03} \tilde{E} + g_{00} \tilde{L}	 \quad , \label{eq:tdotphidot}
\end{align}
where we
again denote
$D = \left ( -g_{00}g_{33} + g_{03}^2 \right )$.
Inserting equation (\ref{eq:tdotphidot}) into equation (\ref{eq:lagrange})
and simplifying yields
\begin{align}
\dot{r}^2 = \frac{1}{g_{11}D} \left(  \tilde{E}^2 g_{33}+ 2g_{03}\tilde{L}\tilde{E}+ g_{00}\tilde{L}^2+D \right ) \label{Grundgleichung} \quad .
\end{align}
This can be rewritten as
\begin{align}
\frac{1}{2}\tilde{E}^2= \frac{1}{2}\dot{r}^2 + V(r,\tilde{E},\tilde{L}) \quad \label{Potentialgleichung}
\end{align}
with 
\begin{align}
V(r,\tilde{E},\tilde{L}) =-  \frac{1}{2g_{11}D} \left( \tilde{E}^2 (g_{33}-Dg_{11})+ 2\tilde{L}\tilde{E}g_{03}+\tilde{L}^2g_{00}+D \right ) \label{eff_potential} \\
= \frac{\tilde{L}^2}{2r^2} -\left(\frac{m}{r}-\frac{B}{4r^3}\right)\left(1+ \frac{\left(\tilde{L}+a\tilde{E}\right)^2}{r^2}\right)+\frac{(1-\tilde{E}^2)a^2}{2r^2}+\frac{1}{2}\label{effpot}
\end{align}
Accordingly the radial motion of a geodesic in the equatorial plane is equivalent to the classical motion of a body with unit mass and
energy $\tilde{E}^2/2$ in a complicated effective potential $V(r,\tilde{E},\tilde{L})$. This concept becomes particularly instructive 
for the Schwarzschild solution. In this case it holds $a=0$, and the effective potential does not depend on $\tilde{E}$:
\begin{align}
\label{eq:pot_schwarzschild}
V_{S}(r,\tilde{L}^2) =\frac{1}{2}-\frac{m}{r}+\frac{\tilde{L}^2}{2r^2}-\frac{m\tilde{L}^2}{r^3}+\frac{B}{4}\left(\frac{1}{r^3}+\frac{\tilde{L}^2}{r^5}\right) \quad .
\end{align}
The terms $-m/r$ and $\tilde{L}^2/r^2$ correspond to the classical
gravitational and centrifugal potential, respectively. In GR the
negative term proportional to $1/r^3$ causes the fall of particles
into the singularity at $r=0$, which is avoided in pc-GR 
due to the repulsive potential proportional to $(1/r^3 + 1/r^5)$. 
If we consider a correction of order $B/r^n$ with $n>3$, in the last term of equation~(\ref{eq:pot_schwarzschild}) the exponents of $r$ in the denominator are $n$ and $n+2$, respectively, and the repelling character of the potential for small $r$ is even stronger. Also, an increasing $B$ does not change the qualitative behavior of the potential.

Whilst for the Kerr metric the effective potential is more complicated and depends not only on $r$ and $\tilde{L}$, but also on $\tilde{E}$, 
it can be used to study the motion along
geodesics. Of particular importance are circular orbits,
which are given by the simultaneous solutions
of $V=\tilde{E}^2/2$ and $\frac{\partial V}{\partial r}=0$.
That is, we consider the set of $r$-dependent functions
$V(r;\tilde{E},\tilde{L})$ with parameter values
$\tilde{E}$ and $\tilde{L}$. We now vary these parameters until
we obtain a curve $V(r;\tilde{E}(r_c),\tilde{L}(r_c))$ such that
this curve takes on a minimum at $r=r_c$
($c$ stands for {\it circular orbit})
and has the value $V(r;\tilde{E}(r_c),\tilde{L}(r_c))=\tilde{E}^2/2$.
The radius $r_c$ together with the parameters $\tilde{E}(r_c)$,
$\tilde{L}(r_c)$ corresponds to a stable circular orbit.
It is convenient to consider the potential $\hat{V}=V-\tilde{E}^2/2$
rather than $V$. The condition for stable circular orbits
is then $\hat{V}=0$ and $\fracpd{\hat{V}}{r}=0$, that is the
function $\hat{V}(r;\tilde{E},\tilde{L})$ has a double root at $r=r_c$.

Instead of trying to solve these conditions directly via equation (\ref{effpot}) and its derivative, we use $\omega\dot{t}=\dot{\varphi}$ as introduced in section \ref{sec:angularfrequency} together with equation (\ref{eq:tdotphidot}). We get
\begin{align}
\tilde{E} &= -\tilde{L} \frac{c\gaa + \omega \gad}{c\gad + \omega\gdd} \quad .
\label{eq:LundE1}
\end{align}
At this point we can use equation (\ref{Grundgleichung}) for geodesic circular orbits ($\dot{r} = 0$) and insert equation (\ref{eq:LundE1}):
\begin{equation}
0 = \tilde{L}^2 \gdd \left ( \frac{c\gaa + \omega \gad}{c\gad + \omega \gdd} \right )^2 - \tilde{L}^2\gad  \frac{c\gaa + \omega \gad}{c\gad + 
\omega \gdd} + \tilde{L}^2 \gaa + D \quad.
\end{equation}
A rather cumbersome rearrangement of this equation yields (together with equation (\ref{eq:LundE1}))
\begin{align}
\tilde{L}^2=\frac{L^2}{\mu^2c^2} &= \frac{\left (c \gad + \omega \gdd \right )^2}{\gdd\omega^2 + 2\gad\omega c+ \gaa c^2} \notag\\
\tilde{E}^2=\frac{E^2}{\mu^2c^4} &= \frac{\left (c \gaa + \omega \gad \right )^2}{\gdd\omega^2 + 2\gad\omega c+ \gaa c^2} \quad .
\label{eq.LundE2}
\end{align}
In these equations $E$ and $L$ are constants of motion, which correspond to energy and angular momentum of a test particle on a 
circular geodesic. 
Both $E$ and $L$ have to be real numbers, and accordingly, the right hand side of equation (\ref{eq.LundE2}) has to be positive. While the numerator is always positive, the denominator corresponds to equation (\ref{eq:bedingungands}) in section~\ref{sec:angularfrequency} and thus can be written as
\begin{align}
\label{eq:EL_real_omega}
\gdd\omega^2 + 2\gad\omega c+ \gaa c^2=g_{33}\left(\omega-\bar{\omega}_{+}\right)\left(\omega-\bar{\omega}_{-}\right)\quad,
\end{align}
where we have used the respective limiting orbital frequencies for general orbits. 
The factor $g_{33}$ is always negative, and thus the product $(\omega-\bar{\omega}_+)(\omega-\bar{\omega}_-)$ has to be negative for $E^{2}$ and $L^{2}$ to be positive. Recall that in pc-GR $\bar{\omega}_->0$ and $\bar{\omega}_+<0$ (see section ~\ref{sec:angularfrequency}). If $\omega>0$, one has $(\omega-\bar{\omega}_+)>0$, and thus it has to hold $\omega<\bar{\omega}_-$ for the expression (\ref{eq:EL_real_omega}) to be positive. For $\omega<0$, the term $(\omega-\bar{\omega}_-)$ is negative, and the expression (\ref{eq:EL_real_omega}) is positive for $\omega>\bar{\omega}_{+}$. It follows, that 
the constraint of positive $E^2$ and $L^2$ is equivalent to the constraint,
that the orbital frequency $\omega$ of a stable circular orbit is within
the limits given by $\bar{\omega}_{\pm}$ derived for general orbits.

We will show now, that for classical GR the conditions $\hat{V}=0$ and  $\fracpd{\hat{V}}{r}=0$ can only be fulfilled for $\tilde{E}^{2}< 1$, see 
\cite{wilkins}.
Let us, for the moment, consider the classical GR, from equation (\ref{effpot}) with $B=0$ and $\hat{V}=V-\tilde{E}^2/2$ one obtains the expression
\begin{align}
\hat{V}(r,\tilde{E},\tilde{L}) 
&= \frac{\tilde{L}^2}{2r^2} -\frac{m}{r} - \frac{m\left(\tilde{L}+a\tilde{E}\right)^2}{r^3} \notag \\
& \quad +\frac{(1-\tilde{E}^2)a^2}{2r^2}+\frac{1}{2}\left(1-\tilde{E}^{2}\right)\quad .
\end{align}
For $r\to 0$ it holds $\hat V\to -\infty$, which has the consequence that if $\hat V$ has a minimum at $r_c$ with $\hat{V}(r_c)=0$, there has to be another root in the interval $(0,r_c)$.

The roots of $\hat{V}$ are identical to the roots of the polynomial
\begin{align}
P(r,\tilde{E},\tilde{L})&=  r^{3}\hat{V}(r,\tilde{E},\tilde{L})\notag \\
&=\frac{1}{2}\left(1-\tilde{E}^{2}\right)r^{3} -mr^{2}  \notag\\
& \quad +  \frac{1}{2}\left(\tilde{L}^{2}+\left(1-\tilde{E}^2\right)a^{2}\right)r -m\left(\tilde{L}+a\tilde{E}\right)^2
\end{align}
According to Descartes' rule of signs the number of positive roots is smaller than or equal to the number of variations in the sign in the polynomial, with multiple roots counted separately, see \cite{anderson}. Given a double root at $r_c$ and another root
in the interval $(0,r_c)$, this demands three changes of sign in
the polynomial $P(r,\tilde{E},\tilde{L})$.
This is only possible if $\tilde{E}^2<1$.
Clearly now,
this is a necessary (but not sufficient) condition for the
occurrence of a stable circular orbit.

We now turn to the case of pc-GR with $B\neq 0$.  From equation (\ref{effpot}) with $\hat{V}=V-\tilde{E}^{2}/2$ we obtain
\begin{align}
\hat{V}(r,\tilde{E},\tilde{L}) 
=& \frac{\tilde{L}^2}{2r^2} -\frac{m}{r} - \frac{m\left(\tilde{L}+a\tilde{E}\right)^2}{r^3}+\frac{(1-\tilde{E}^2)a^2}{2r^2}\notag \\
&+\frac{1}{2}\left(1-\tilde{E}^{2}\right) + \frac{B}{4r^3} + \frac{B\left( \tilde{L}+a\tilde{E}\right)^{2}}{4r^{5}} \quad .
\end{align}
For $r\to 0$ it holds $\hat{V}\to \infty$ (repulsive potential),
whereas for $r\to \infty$ it
approaches
$\frac{1}{2}\left(1-\tilde{E}^{2}\right)$ from below. It follows that if we have a minimum of $\hat{V}$ at $r_c$ with $\hat{V}(r_c)=0$, for $\tilde{E}^2\geq 1$ there is another positive root at some value $r'>r_c$. In this case and as well for $\tilde{E}^2<1$, there might be additional roots, but their existence is not a necessary condition for a double root which is a minimum at $r_c$.\\

The roots of $\hat{V}$ are identical to the roots of the polynomial

\begin{align}
P(r,\tilde{E},\tilde{L})= & r^{5}\hat{V}(r,\tilde{E},\tilde{L})\notag \\
=&\frac{1}{2}\left(1-\tilde{E}^{2}\right)r^{5} -mr^{4} +  \frac{1}{2}\left(\tilde{L}^{2}+\left(1-\tilde{E}^2\right)a^{2}\right)r^3 \notag\\
&+\left[\frac{B}{4}-m\left(\tilde{L}+a\tilde{E}\right)^2\right]r^2 + B\left( \tilde{L}+a\tilde{E}\right)^{2} \label{ppc}
\end{align}
For $\tilde{E}^2<1$ there are always two changes of sign in the first
three terms. Thus,
the necessary condition for the existence of a stable
circular orbit is always fulfilled. For $\tilde{E}^2>1$,
the polynomial has to show three changes of sign (recall that the
double root at the minimum is counted separately).
Therefore, the two conditions
\begin{align}
\label{eq:e_constraint}
\tilde{L}^{2}+\left(1-\tilde{E}^2\right)a^{2} &>0  \notag \\
\frac{B}{4}-m\left(\tilde{L}+a\tilde{E}\right)^2 &<0 \quad .
\end{align}
 have to hold.
It therefore follows, that in pc-GR it is not the general requirement $\tilde{E}^2<1$, but the weaker restriction (\ref{eq:e_constraint}) for $\tilde{E}^{2}>1$ that is a necessary condition for the occurrence of stable circular orbits.

There might be various combinations $r_c,\tilde{E}(r_c),\tilde{L}(r_c)$ corresponding to stable circular orbits. The last stable orbit can be found by setting $ \fracppd{V}{r}|_{r_c}=0$. Together with equations (\ref{eq.LundE2}) and (\ref{eff_potential}) this is equivalent to
\begin{align}
& \gdd'' (\gaa + \omega \gad)^2 + \gaa'' (\gad +\omega \gdd)^2 \notag \\
&- 2 \gad'' (\gaa +\omega \gad )(\gad +\omega \gdd) \notag \\
&+ D''(\omega^2 \gdd +2\omega \gad + \gaa) = 0 \quad. 
\label{letztestabilebahn}
\end{align}
Unfortunately equation (\ref{letztestabilebahn}) has a quite complicated form when we insert the pc-Kerr metric [done with
\cite{mathematica}]
\begin{align}
&\frac{1}{r \left(4 r^5+a^2 \left(3 B-4 m r^2\right)\right)^2}\left[-6 a^4 \left(15 B^2 r-32 B m r^3+16 m^2 r^5\right) \right. \notag \\
&\pm 4 a r^{5/2} \sqrt{-3 B+4 m r^2} \left(15 B^2+24 m r^4 (2 m+r)-10 B r^2 (4 m+3 r)\right) \notag \\
&+4 r^5 \left(-15 B^2+8 m r^4 (-6 m+r)+2 B r^2 (20 m+3 r)\right) \notag \\
&+a^2 \left(45 B^3-18 B^2 r^2 (10 m+17 r)-96 m r^6 \left(2 m^2+5 m r+r^2\right) \right. \notag \\
& \left.+8 B r^4 \left(38 m^2+96 m r+15 r^2\right)\right) \notag \\ 
&\pm 8 a^3 \left(4 m r^{9/2} (4 m+3 r)\sqrt{-3 B+4 m r^2} \right.\notag \\
&\left. \left. -3 B r^{5/2} (8 m+5 r) \sqrt{-3 B+4 m r^2}+9 B^2 \sqrt{-3 B r+4 m r^3}\right)\right] = 0 \label{eq:letztebahnausgeschrieben} \quad .
\end{align}
For $B=0$ this equation greatly simplifies to [done with \cite{mathematica}]
\begin{equation}
r^2 - 6mr \pm 8a \sqrt{mr} -3a^2 = 0 \quad , \label{eq:letztebahnklassisch}
\end{equation}
which is the same as in \cite{bardeen}. The upper sign here refers to a
co-rotating object. Up to now we have not found analytical solutions
for equation (\ref{eq:letztebahnausgeschrieben}). Numerical investigations
showed, that equation (\ref{eq:letztebahnausgeschrieben}) has two solutions for
counter-rotating objects (see Fig. \ref{fig:retro}).
\begin{figure}	
\begin{center}
\includegraphics[width=\columnwidth,]{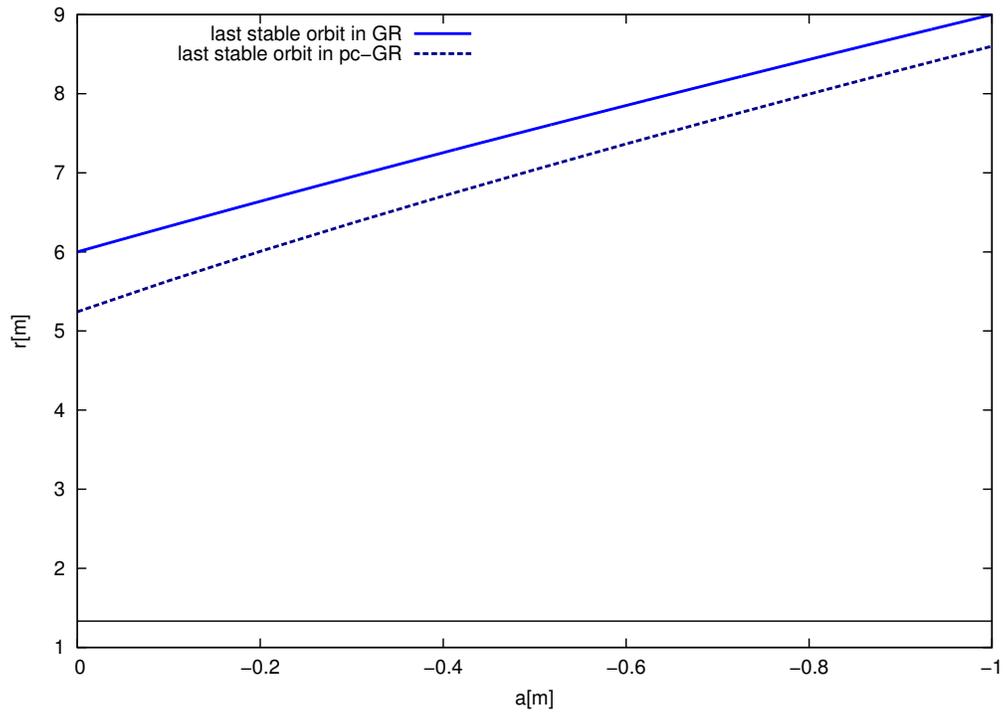} 
\begin{flushleft}
\caption{Critical stable orbits for retrograde test masses
\label{fig:retro} - the parameter $B$ is set to $ B = \frac{64}{27} m^3$.
}
\end{flushleft}
\end{center} 
\end{figure}   
However only the outer one
of these solutions is physically relevant as one has to 
consider the constraints given by equation (\ref{eq:omegaconstraint}).
The origin of the complex structure of the last equation is
the addition of the $B/(2r^3)$ term in the metric, resulting in
a new rich structure of solutions.

 Equation (\ref{eq:letztebahnausgeschrieben}) also has two
solutions for co-rotating objects but only up to a value of $a > -0.416m$
(see Fig. \ref{fig:pro}).
 For higher absolute values of $a$,
 equation (\ref{eq:letztebahnausgeschrieben}) has no positive real roots for
co-rotating objects. For this case, the second derivative of the potential is always positive,
i.e., {\it a stable orbit always exists}.

As we encounter a new physical phenomenon, namely the existence of two
critical orbits compared to only one in the classical
Kerr metric, we will investigate Fig. \ref{fig:pro}  a little bit further.

 To do so
we divide the illustrated parameter space into four different regions: 
\begin{itemize}
 \item[I] This region is bounded 
by the solid line which represents the last stable orbit in GR. For all combinations of $a$ and $r$ above this line $\fracppd{V}{r}$ is positive for both GR and pc-GR. Thus, in this region there exist stable circular orbits in both theories.
\item[II] In this area $ \fracppd{V}{r} $ is positive for pc-GR, in contrast to standard GR where $ \fracppd{V}{r}$ is negative. This means that for all combinations of $r$ and $a$ in this dark shaded area there are stable circular orbits in pc-GR but not in GR.
\item[III] Here, between the dashed and dash-dotted curves, $\fracppd{V}{r}$ is negative for all values of $r$ and $a$ even for pc-GR. So there are no stable geodesic circular orbits in this light shaded area regardless of the theory concerned. 
\item[IV] In principle, orbits would be stable here for pc-GR, $ \fracppd{V}{r} $ is positive in this region, but the constraint (\ref{eq:omegaconstraint}) excludes this area for general orbits (i.e. $ds^2$ is negative in this area).
\end{itemize}

The most interesting area now is II as we do have a different behavior
for pc-GR compared to GR. In this region there exist stable orbits in
pc-GR but not in GR. Especially for values of $a < -0.416m$ we have $
\fracppd{V}{r} > 0$ for all values of $r$ which means that here no
last stable orbit occurs but rather all orbits are stable in
pc-GR. Going to values of $a > -0.416$ we also see a new
phenomenon. In contrast to GR there now exists an area of unstable
orbits which has an upper (dashed curve - corresponds to the solid
curve in GR) and lower (dash-dotted curve - not existent in GR)
bound. Thus, in pc-GR there is a ring-shaped area of unstable orbits
with a stable zone in the middle around the central mass while
  in GR this area is a disc extending to the surface of the black
  hole with no stable zone below the last stable orbit.

For radii smaller than $\frac{4}{3}m$ the solutions of equation (\ref{Kreisfrequenz_neu}) become imaginary. However, $r=\frac{4}{3}m$  is the
position of the global minimum of $g_{00}$ (and therefore the effective potential), describing the final radius of a large mass.
For a description at smaller radial distances, the distribution of the
object's mass has to be included.

In summary, the physical behavior in pc-GR for counter-rotating orbits is not 
very different to the standard GR except that the last stable orbit
is a little closer to the central mass. For co-rotating orbits 
however we get a new physical behavior.

\begin{figure*}
\vbox to220mm{\vfil 
\begin{center}
\includegraphics[width=\textwidth]{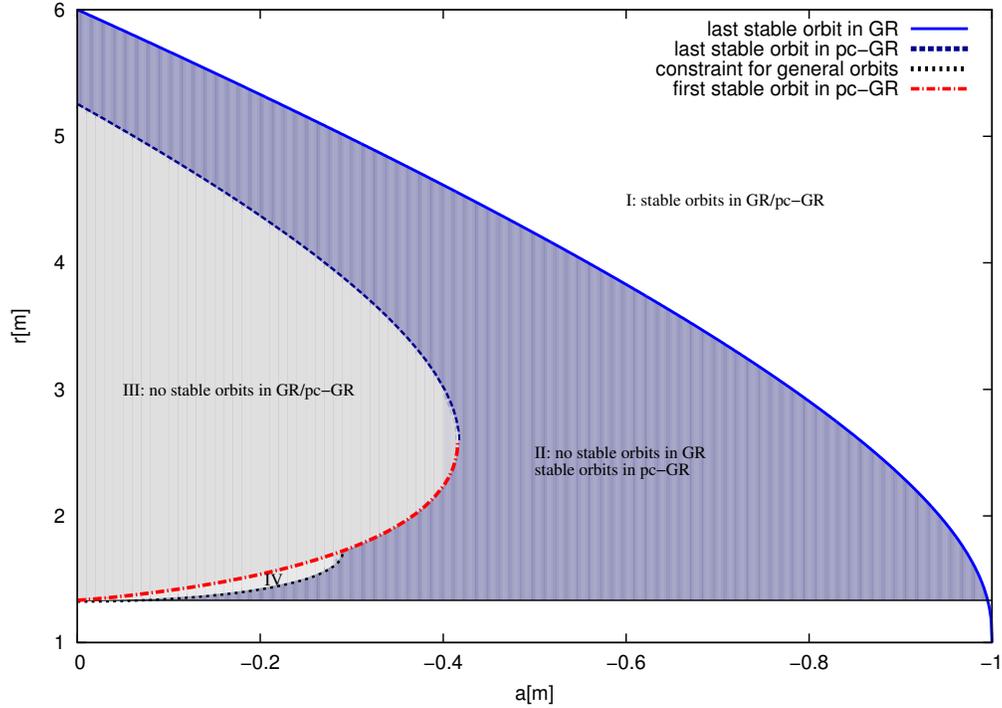} 
\begin{flushleft}
\caption{Critical stable orbits for prograde test masses
\label{fig:pro}
(The curves describe from top to bottom: 1. The last stable orbit
in standard GR (thick solid), 2. the `last' stable orbit in pc-GR (dashed), 3. the `first' stable
orbit in pc-GR (dash-dotted), 4. the limit to general orbits given by equation (\ref{eq:omegaconstraint}) 
(dotted) and 5. the point where the pc-equations become imaginary ($r= \frac{4}{3}$, thin solid)). We now have to distinguish between four different areas. In the unshaded area (I) orbits are stable both in GR and pc-GR, whereas in the dark shaded area (II) orbits are only stable in pc-GR. Both lighter
shaded areas (III and IV) do not contain stable orbits at all. The plot is done for a value of $B = \frac{64}{27}m^3$.
}
\end{flushleft}
\end{center} 
\vfil}
\label{landfig}
\end{figure*}

\clearpage

\section{Conclusions and outlook}

We have identified several observables and proposed associated
 possibilities
for astronomical measurements, which may distinguish between
different predictions of pseudo-complex General Relativity and standard GR. 
For distances comparable to the Schwarzschild radius
we showed that the angular frequencies and allowed orbits of particles orbiting around a dense central object 
(`gray star' in pc-GR or black hole in GR)  and the gravitational redshift differ
significantly between GR and pc-GR.
We have illlustrated that
the observables related to the orbital frequency may be
accessible by measuring plasma clouds or flares from the
vicinity of a gray star/black hole.
We also showed how techniques to determine the gravitational redshift
close to the central object could be applied.

These are our new findings:
For particles moving in the
equatorial plane the angular frequencies for pro- and retrograde geodesic orbits in pc-GR metrics are smaller than those 
for particles moving
according to GR.
 We found that
the limits of the angular frequency for general circular motions in the equatorial
plane to have a wider range in pc-GR than in GR leading to a weaker
frame dragging effect. Therefore, we propose
to measure plasma clouds or flares close to the Galactic center
or M87. We
elaborated the differences between pc-GR and GR considering the redshift
of an emitting particle at rest.
We found that the redshift in pc-GR is smaller
than in GR for central
objects with the same mass and density. Especially the poles of rotating dense objects are much brighter,
so that the inclination has a
pronounced effect on the blackness of the central star.
To observe this,
we proposed to directly
investigate the optical appearance of the
dense object preferably by radio astronomy or
 by using
spectral methods e.g. tracing of X-ray iron lines.
 Finally, we considered the effective potentials of the GR and pc-GR geometries and their
impact on the allowed circular orbits.
By deriving the effective potentials and introducing the
requirements for stability of circular orbits,
 we discussed the restraints of the angular momentum and the energy, so that we
could identify the stable orbits in pc-GR and GR.
We showed,
that in pc-GR the concept of one last stable orbit 
is changed. Depending on the
momentum of the central mass there is a last stable orbit (a = 0)
and it exists a forbidden zone between a `last' and a
`first' stable orbit (at a certain point below the `last' stable orbit orbits get stable again; 0 $>$ a $>$ -0.416m) or
for all radii a stable orbit can be found ( a $<$ -0.416m)!\\

Here, we have shown how the pseudo-complex General Relativity changes the standard picture.
We presented various predictions of pc-GR in the vicinity of black-hole candidates which are 
testable with astronomical methods. 
The field of pc-GR is rather new, hence there are still many open questions. Among them are 
other aspects yet to be investigated like the significance of pc-GR at a cosmological length 
scale as well as its implications at a microscopic scale.

\section*{Acknowledgments}

The authors express sincere gratitude for the possibility to work at
the {\it Frankfurt Institute of Advanced Studies} and at the
{\it GSI} with their excellent
working atmosphere.
Peter Otto Hess also acknowledges financial support from FIAS, DGAPA-PAPIIT
(IN103212), DGAPA and CONACyT. Gunther Caspar acknowledges financial support from {\it Frankfurt Institute for Advanced Studies}.
Thomas Sch\"onenbach acknowledges support from {\it Stiftung Polytechnische Gesellschaft Frankfurt am Main}.  
Andreas M\"uller acknowledges support by the DFG cluster of excellence 'Origin and Structure of the Universe'.
Thomas Boller is grateful to Stefan Gillessen for intensive
discussion on the Galactic Center research lead by the infrared group of the
Max-Planck-Institute for extraterrestrial physics. 
The authors thank Tom Dwelly for critical reading of the paper.

\end{document}